\documentclass[twocolumn,showpacs,prbprintnumbers,prb,fleqn,superscriptaddress]{revtex4}
\usepackage{epsfig}
\usepackage{graphicx}
\usepackage{amsfonts}
\usepackage[dvips]{color}

\newcommand{\ct}{\cite}
\newcommand{\bi}{\bibitem}
\newcommand{\be}{\begin{equation}}
\newcommand{\ee}{\end{equation}}
\newcommand{\ba}{\begin{eqnarray}}
\newcommand{\ea}{\end{eqnarray}}

\begin{document}
\title{Path dependent scaling of geometric phase near a quantum multi-critical point}
\author{Ayoti Patra}
\email{ayoti@iitk.ac.in}
\author{Victor Mukherjee}
\email{victor@iitk.ac.in}
\author{Amit Dutta}
\email{dutta@iitk.ac.in}
\affiliation{Department of Physics, Indian Institute of Technology Kanpur,
Kanpur 208 016, India} 

\begin{center}
\begin{abstract}
We study the geometric phase of the ground state in a  one-dimensional transverse $XY$ spin chain in the vicinity of a quantum multi-critical point.
 We approach the multi-critical point along different paths  and estimate the geometric phase by applying a rotation in all spins about $z$-axis by an angle $\eta$.
Although the geometric phase itself vanishes at the multi-critical point, the derivative  with respect to the anisotropy parameter of the model  shows  peaks
 at different points on the ferromagnetic side close to it 
 where the energy gap is a local minimum; we call these points `quasi-critical'. The value of the derivative at any quasi-critical point scales
 with the system size in a power-law fashion with the exponent varying continuously with the parameter $\alpha$ that defines a path, upto a critical value 
$\alpha = \alpha_{c}=2$. 
For $\alpha > \alpha_{c}$, or on the paramagnetic side no  such peak is observed. Numerically obtained results are in perfect agreement with analytical predictions.
\end{abstract}
\end{center}

\maketitle

\section{Introduction}
 Quantum Phase Transitions (QPT) in quantum many body systems  are being studied extensively in recent years\ct{sachdev99, chakrabarti96,continentino}. 
The possibility of experimental studies on ultracold atoms trapped in optical lattices, which can for e.g., undergo a  mott insulator to a superfluid transition, 
have opened new 
avenues to investigate quantum phase transitions\ct{greiner02, cirac98}.  QPTs occur at absolute zero temperature and are associated with a fundamental change in 
the ground state 
of the system\ct{sachdev99}.  A quantum critical point is characterized by a diverging length scale as well as a diverging time scale, namely the relaxation time 
of the
quantum system. This characteristic time scale  is the inverse of the minimum  energy gap of the underlying quantum Hamiltonian which vanishes at the quantum 
critical point. Recently,
 QPTs have also been  studied from the viewpoint of quantum information theory. Quantities like concurrence \ct{Osterloh02}, entanglement 
entropy\ct{vidal03, kitaev06}, 
fidelity susceptibility\ct{zanardi06, venuti07, giorda07,  gurev08, shigu08, zhou08,schwandt09,grandi10, polkovnikovrmp} and geometric 
phases\ct{carollo05,zhu06,hamma06,zhu08, cui06}  have been studied in the vicinity of quantum critical points, and are found to capture the singularities associated with them. 
A close relation between geometric phase(GP) and magnetic susceptibility has also been established for transverse XY chain  for QPTs driven by 
transverse magnetic field \ct{quan09}. It is worth mentioning that the ground state GP in a Heisenberg XY Model has been experimentally observed very recently using NMR interferometry \ct{peng10}. 

In the  present work, we investigate the scaling  behavior of GP \ct{panchratnam56, berry84} near a quantum multi-critical point (MCP) generated by the application of rotation
 on each spin of  a quantum spin chain around $z$-axis. Berry showed in that in addition  to the usual
 dynamic phase,  a phase purely of geometric nature is  accumulated on the wave function of a quantum system under an adiabatic and cyclic change of the Hamiltonian\ct{berry84}. Quantum critical points are regions of vanishing energy gaps, and consequently are accompanied by 
non-analyticities in various observables of the system. In a quantum critical system, the GP  of the ground state of the system, which depends on the energy gap, can capture the associated singularities\ct{zhu06, carollo05, hamma06}. In fact, GP can also be related to the imaginary part of a general geometric tensor, 
whose real part on the other hand gives the fidelity susceptibility which is the rate of change of the ground state of the Hamiltonian following an infinitesimal change
in its parameters\ct{venuti07,giorda07, basu10,jun10}. 

We  exploit the integrability of  spin-1/2 $XY$ chain \ct{lieb61,barouch70,kogut79,bunder99}  
 to investigate the nature of the geometric phase close to the MCP. 
 Recently, it has been shown that a GP difference  between the ground and the first excited state exists in an isotropic $XY$ chain (i.e., the $XX$ chain) 
if and only if the closed evolution path circulates a region of 
criticality \ct{carollo05}. In this paper,  we use a dynamical scheme so that the MCP is approached along various paths characterized by a parameter $\alpha$. 
The non-contractible GP of the ground state is studied following the behavior of GP and its derivative close to 
the Ising transition point of an anisotropic $XY$ chain\ct{zhu06}.  In the case of a MCP, we find that the derivative of the GP  with  the anisotropy parameter scales with the chain length
 in a non-trivial 
fashion with an exponent depending on the parameter $\alpha$.  Moreover, it does not peak right at the MCP, rather it peaks at points close to the MCP on the 
ferromagnetic side 
(which are so-called `quasi-critical points' \ct{deng09, dutta10}), where the energy gap is a local minima. However, beyond a limiting value of
 $\alpha = \alpha_c$ no such peak is 
observed and GP as well as its derivative at the MCP is zero for all values of $\alpha$.  

\section{The model and the geometric phase.}
\begin{figure}[ht]
\begin{center}
\includegraphics[width=7.9cm]{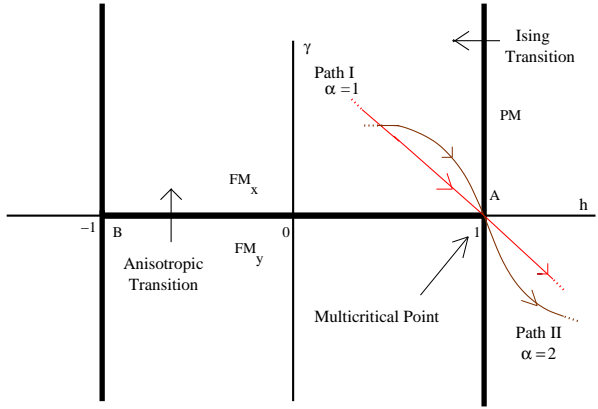}
\end{center}
\caption{The phase diagram of a one-dimensional $XY$ model in a transverse field. The vertical bold lines  ($h= \pm 1$) denote Ising transitions 
from the ferromagnetic 
to the paramagnetic phase. The horizontal bold line stands for the anisotropic phase transition from a ferromagnetic phase with magnetic ordering 
in $x$-direction to a ferromagnetic phase with 
ordering in $y$-direction.The multi-critical points are at A ($h=1$, $\gamma=0$) and B ($h=-1$, $\gamma=0$).
 We show different paths of approaching the MCP A corresponding to different values of $\alpha$; path I (path II) is for $\alpha=1$ 
($\alpha=2$). We show that for $\alpha < 2$, we observe quasi-critical peaks  and there is a continuously varying effective scaling exponent 
for $d\beta _{g}/d\gamma$
with the chain length, while for $\alpha > 2$, no peak is observed. }
\label{Fig:pd}
\end{figure}

The model we consider is a one-dimensional spin-1/2 $XY$ model in a transverse field with nearest neighbor ferromagnetic interactions  given by the Hamiltonian\ct{lieb61, barouch70,bunder99, duttarmp}
\be
H = - \frac{1}{2} ~\sum_n ~[(1+\gamma) \sigma^x_n \sigma^x_{n+1} + (1- \gamma) \sigma^y_n \sigma^y_{n+1} + h \sigma^z_n],
\label{h1} 
\ee 
where $\sigma$'s are Pauli spin matrices satisfying the usual commutation relations. The parameter $h$ is the magnetic field applied in the $z-$direction and $\gamma$ 
measures the anisotropy in the in-plane interactions.  
The phase diagram of the system, plotted in $h - \gamma$ plane is shown in Fig.~(\ref{Fig:pd}) where the vertical bold lines at $h=\pm 1$ represent quantum transitions
 from the ferromagnetic to the paramagnetic phases that belong to the transverse Ising universality class and hence called the `Ising transitions'.
 The horizontal bold line  at $\gamma = 0$ with the transverse field lying  between $1$ and $-1$ represents transitions between ferromagnetic ordered phases with ordering 
in $x$ and $y$ directions, respectively \ct{ barouch70, bunder99}. The points A and B ($h= \pm 1$ at $\gamma=0$, respectively)
 where Ising and anisotropic lines meet are the multi-critical points \ct{damle96}  which are our subject of interest.
Analyzing the energy spectrum of the Hamiltonian (\ref{h1}) \ct{bunder99}, it can be easily shown that the energy gap scales with the momentum $k$ as $k^z(=k)$ at any Ising 
critical point ($h=\pm 1, \gamma \neq 0$) so that  the dynamical exponent $z=1$. On the other hand, for the critical mode the gap scales as $|h-1|^{\nu z}(= |h-1|)$ yielding the 
correlation length exponent $\nu=1$. Similarly, one finds for MCP  that $z=2$ and $\nu =1/2$ \ct{damle96}; $\nu z=1$ in either cases.

We propose a dynamical scheme that enables us to approach the MCP along the path defined by\ct{dutta10}
\begin{equation}
h(\gamma) = 1-|\gamma |^{\alpha} sgn(\gamma), 
\label{path}
\end{equation}
so that the system hits the MCP A at $h=1,\gamma=0$ (see Fig.~(\ref{Fig:pd})). Close to a MCP, the path of approach plays an important
role as also to be shown below in the scaling of any dynamical response; the  dynamical scheme (\ref{path}) in which the path can be changed through tuning
the parameter $\alpha$  therefore facilitates the study on the path-dependent scaling of GP close to the MCP.

In order to investigate the GP in this system, we introduce a new family of Hamiltonians that can be described by applying a rotation 
of $\eta$ around the $z$ direction to each spin \ct{carollo05, zhu06},
i.e., $H_{\eta} = g_{\eta}Hg_{\eta}^{\dagger}$ with $g_{\eta} = \prod_{n} \exp(i \eta \sigma^z_n/2)$. 
For the family of Hamiltonians thus generated, the energy spectrum remains the same leaving the critical behavior unaltered. 
The Hamiltonian $H(\eta)$ can be diagonalized by using the Jordan-Wigner transformation which transforms
the spin operators into fermionic operators $a_n$ and $a_n^{\dagger}$ via the relations\ct{jordan28,lieb61}
 $a_n = (\prod _{l<n} \sigma_l^z)\sigma_n^{\dagger}$. Employing a Fourier transformation 
$d_k = \sum_j \left[a_j \exp(-i2 \pi jk/N)\right]/\sqrt{N} $ followed by Bogoliubov transformation, one can recast
 Hamiltonian (\ref{h1}) to the diagonal form
 $H = \sum_k \Lambda_k(c^{\dagger}_k c_k -1)$, 
where $\Lambda_k = \sqrt{(h + \cos(2 \pi k/N))^2 + \gamma^2 \sin ^2(2 \pi k/N)}$, $ c_k = d_k \cos \frac{\theta_k}{2} -d_{-k}^{\dagger} e^{2i \eta} \sin \frac{\theta_k}{2}$, 
and the angle $\theta_k$ is given by $\cos \theta_k = (\cos \frac{2 \pi k}{N} + h)/ \Lambda_k$.

The ground state is a tensor product of states, each lying in the two-dimensional Hilbert space spanned by
 $|0\rangle_k|0\rangle_{-k}$ and $|1\rangle_k|1\rangle_{-k}$, where $|0\rangle_k$ and $|1\rangle_k$ are the vacuum and the excited state for $k$-th mode
$|1\rangle_k|1\rangle_{-k}= d^{\dagger}_k d^{\dagger}_{-k} |0\rangle_k |0\rangle_{-k}$. Under the unitary transformation (i.e. rotation about $z-$axis), the ground state of the 
transformed Hamiltonian  
picks up an additional phase factor and is given by
 $|g \rangle = \prod_{k} (\cos \frac{\theta_k}{2} |0\rangle_k|0\rangle_{-k} - ie^{2i \eta} \sin \frac{\theta_k}{2}|1\rangle_k|1\rangle_{-k})$. 
 The GP of the ground state, accumulated by varying the angle
 $\eta$ from $0$ to $\pi$, is described by\ct{berry84} $\beta_g = - i/N \int_0^{\pi} \langle g|\partial_{\eta}|g\rangle d \eta $ , and is given by 
\begin{equation}
\beta_g = \frac{\pi}{N} \sum_{k} (1-\cos \theta_k).
\label{beta2}
\end{equation}
This equation in the thermodynamic limit ($N \rightarrow \infty$), where the critical properties are studied , is given by 
\begin{equation}
\beta_g = \int_0^{\pi} (1-\cos \theta_{\phi}) d \phi ,
\label{h2}
\end{equation}
where the summation $1/N\sum_{k}$ is replaced by the integral $1/\pi\int_0^{\phi}$ with $\phi =2 \pi k/N$, so that 

\ba
\cos \theta_{\phi} &=& (\cos \phi + h)/ \Lambda_{\phi} ~~~{\rm and}~~~ \nonumber\\
\Lambda_{\phi} &=& \sqrt{(h + \cos \phi)^2 + \gamma^2 \sin^2 \phi}. 
\label{costheta}
\ea
Carollo and Pachos\ct{carollo05} studied the behavior of the GP close to the anisotropic critical point and showed that a non-contractible 
geometric phase difference between the ground state and the first excited state exists when the Hamiltonian encounters a critical point while
 passing through an adiabatic cycle. In a subsequent work, Zhu\ct{zhu06} showed that the  derivative $d \beta_g/d h$ as $h$ is varied shows a peak right at the Ising critical point
 and diverges  logarithmically with the chain length. From the scaling relations
 $d\beta _{g}/dh \sim \kappa_1 \ln N + C_1 $ at the critical point $h=h_c=1$, and $d\beta _{g}/dh \sim \kappa_2 \ln |h -h_c|  + 
C_2 $  for an infinite system (where $C_1$ and $C_2$ are non-universal constants), the critical exponent $\nu$ can be  obtained  from the relation $\nu = \kappa_2/\kappa_1 = 1$.
\section{Results}

\begin{figure}[ht]
\begin{center}
\includegraphics[width=8.1cm]{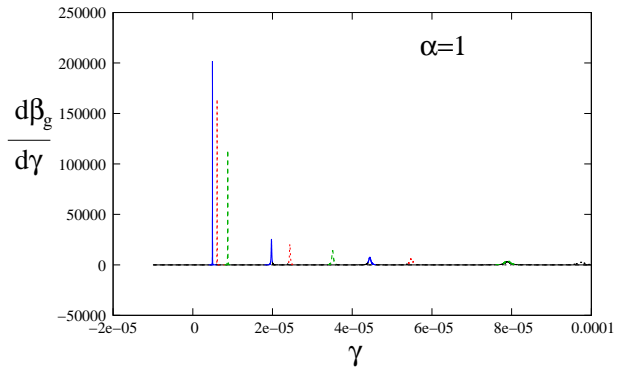}
\end{center}
\caption{(Color Online) The numerically observed variation of  $d\beta _{g}/d\gamma$ with respect to  $\gamma$ for $\alpha = 1$ for different chain lengths, 
$N=2000$ (blue solid line), $N=1800$ (red dotted line)  and  $N=1500$ (green dashed line). In all cases, a series of peaks is observed on the 
ferromagnetic side $(\gamma >0, |h|<1)$ along the path of approach. With increasing $N$ 
the peak height increases, the peaks shift closer to MCP, and the distance between subsequent peaks decreases.}
\label{Fig:qcps}
\end{figure}

\begin{figure}[ht]
\begin{center}
\includegraphics[width=7.9cm]{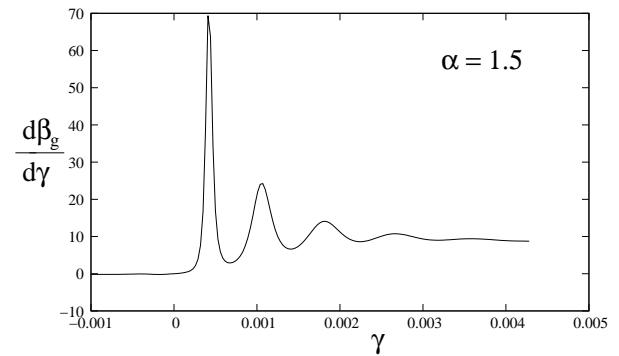}
\end{center}
\caption{A similar behavior as in Fig.~(\ref{Fig:qcps}) for $d\beta _{g}/d\gamma$ as a function of $\gamma$ is observed for $\alpha = 1.5$.}
\label{Fig:qcpnqcp}
\end{figure}

\begin{figure}[ht]
\begin{center}
\includegraphics[width=7.9cm]{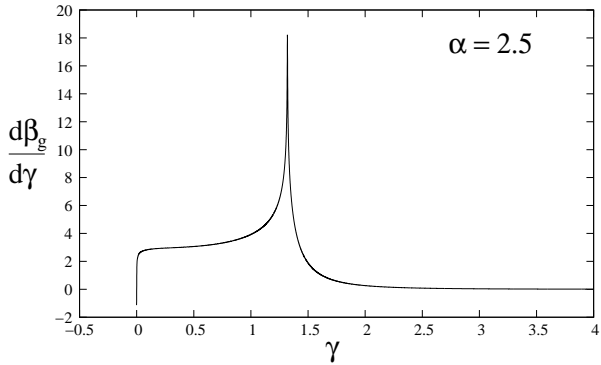}
\end{center}
\caption{Numerically no peak is observed in  $d\beta _{g}/d\gamma$ for $\alpha = 2.5$ which is due to the absence of quasi-critical
points along the path. However,  a  peak is observed corresponding to the Ising critical point at $\gamma=1.319, h=-1$.}
\label{Fig:qcpnqcp2}
\end{figure}

\begin{figure}[ht]
\begin{center}
\includegraphics[width=7.9cm]{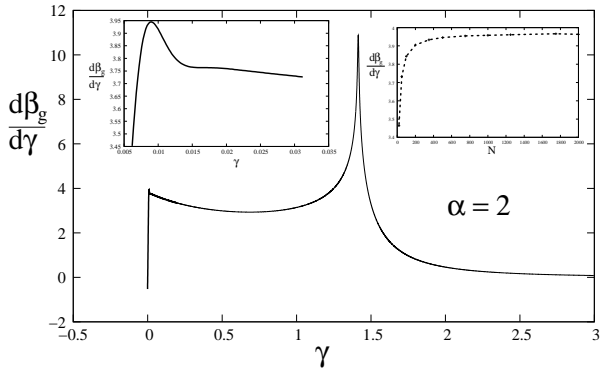}
\end{center}
\caption{ The numerical plot of $d\beta _{g}/d\gamma$ with $\gamma$ for the marginal case $\alpha=2$ indicating a peak close to the MCP (left inset) with 
the peak height
saturating to a constant value with $N$ (right inset) as expected from the analytical scaling relation (\ref{scalingGPN}). The Ising peak is observed at
 $\gamma = 1.414$ where $h=-1$.}
\label{Fig:a2}
\end{figure}

We study the behavior of GP in the vicinity of the MCP `A' (see Fig.~(\ref{Fig:pd})) using the dynamical path given in Eq.~(\ref{path}). 
Let us first consider the case of $\alpha=1$, i.e., a linear path approaching the MCP and estimate the geometric phase and its derivative using
numerical techniques. We observe  a series of peaks in 
the derivative $d \beta_g/d \gamma$ when plotted against $\gamma$ close to the MCP on the ferromagnetic side ($\gamma >0$, $|h|<1$ for the path (\ref{path})) of 
it as shown in Fig.~(\ref{Fig:qcps}). A similar behavior is 
observed for all values of $\alpha < 2$ (see Fig.~(\ref{Fig:qcpnqcp})). This is in contrast to the behavior near an Ising critical point where 
there is only one peak right at the quantum critical point \ct{zhu06} (see Fig.~(\ref{Fig:qcpnqcp2})). 
No such peak is observed on the paramagnetic side ($\gamma <0$, $|h|>1$)  and the GP
 itself becomes trivial at the MCP as seen from Eq.~(\ref{beta2}).  On the other hand, for $\alpha >2$ no peaks are observed close to the MCP (see Fig.~(\ref{Fig:qcpnqcp2})); 
and the case $\alpha=2$ shows a limiting
behavior to be discussed later (see Fig.~(\ref{Fig:a2})).  The appearance of the series
 of peaks  in  $d \beta_g/d \gamma$  can be attributed to the existence of quasi-critical points where the energy spectrum attains a 
local minima, to be explained below.

The scaling behavior of the derivative $d \beta_g/d \gamma$ with the system size for a generic $\alpha$  can be derived in the following way. 
From Eq.~(\ref{path}) we get the spectrum as 
\begin{equation}
\Lambda_{\phi} = \sqrt{ [\cos \phi + (1- |\gamma|^{\alpha} sgn(\gamma))]^2 + \gamma^2 \sin^2 \phi },
\label{spectrum}
\end{equation}
so that using  Eq.~(\ref{h2})  for positive $\gamma$, $d \beta_g/d \gamma$ can be expressed as
\ba
\frac{d \beta_g}{d \gamma} &=& - \int_0^{\pi} \left( \frac{d \cos \theta_{\phi}}{d \gamma} \right) d \phi \nonumber\\
& =& \int_0^{\pi} \left( \frac{\alpha \gamma ^{\alpha-1}}{\Lambda_{\phi}} - 
\frac{A}{\Lambda_{\phi}^3} \right) ~ d \phi,
\label{dvtv}
\ea
where
 $$A = \{\cos \phi + (1- \gamma^{\alpha})\} \{ [\cos \phi + (1- \gamma^{\alpha})] \alpha \gamma^{\alpha -1} - \gamma \sin^2 \phi \}.$$
The first term of the integrand dominates over the second term which can be dropped for further calculations. For $\phi \simeq \pi$, the minima in 
 $\Lambda_{\phi}$ occurs at points where $\gamma^{\alpha} \sim \phi^2$ (shown below) so that
\ba
\frac{d \beta_g}{d \gamma} = \int_0^{\pi} \frac{\alpha \phi^{\frac{2}{\alpha}(\alpha -1)}}{\phi^{(\frac{1}{2})(\frac{4}{\alpha} +2)}}d\phi ~\propto ~ \phi^{2- \frac{4}{\alpha}},
\label{dscaling}
\ea
with $\phi \sim 1/{N}$. Hence we get the scaling relation
\begin{equation}
\frac{d \beta_g}{d \gamma} \propto N^{\frac{4}{\alpha} - 2}.
\label{scalingGPN}
\end{equation}
\begin{figure}[ht]
\begin{center}
\includegraphics[width=7.9cm]{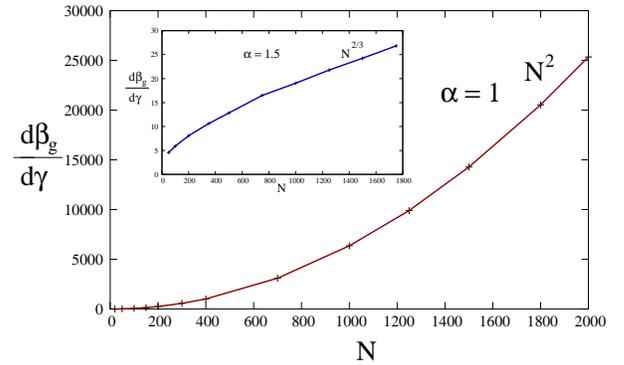}
\end{center}
\caption{Scaling behavior of $d \beta_g/d \gamma$ with $N$ is shown for $\alpha = 1$ and $\alpha=1.5$ (inset). The numerical results perfectly match
with the scaling relation (\ref{scalingGPN}).}
\label{Fig:scale}
\end{figure}

The scaling relation (\ref{scalingGPN}) shows that $d \beta_g/d \gamma$ diverges with the system size for $\alpha <2$ while
 for $\alpha=2$ a saturation is expected. The scaling exponent varies continuously with the path, i.e.,  with the parameter $\alpha$ upto the limiting value
  $\alpha=2$. Fig.~(\ref{Fig:scale}) shows the scaling of $d \beta_g/d \gamma$ with $N$ for $\alpha=1$ and $1.5$ confirming the
analytical scaling (\ref{scalingGPN}) while no peak is expected for $\alpha>2$ as numerically observed in Fig.~(\ref{Fig:qcpnqcp2}). 
With increasing $\alpha$, the divergences seen in Eq.~(\ref{scalingGPN}) is expected to be slower explaining observations in Figs. (\ref{Fig:qcps}) and (\ref{Fig:qcpnqcp}).
 
 Let us now focus on 
 the energy spectrum (\ref{spectrum}); 
  the energy gap is not minimum at the MCP for $\alpha<2$, rather it is minimum at the `quasi-critical' points determined by
 the condition $ \cos \phi +(1-|\gamma|^{\alpha} sgn(\gamma))=0$ or $|\gamma|^{\alpha} sgn(\gamma)=\phi^2$ if we expand $\phi \simeq \pi $ and rescale $\phi \rightarrow \pi - \phi$.
     The derivative of GP shows a peak whenever the energy gap is minimum, i.e., the spectrum hits a `quasi-critical' point 
     resulting in a series of peaks on the ferromagnetic side of MCP as numerically observed in Fig.~(\ref{Fig:qcps}) 
     and Fig.~(\ref{Fig:qcpnqcp}). The condition $|\gamma|^{\alpha} sgn(\gamma)=\phi^2$ does also imply that for increasing chain length, the quasi-critical points 
(and hence the peaks in $d \beta_g/d \gamma$) become closely spaced.
Note that strictly for $N \to \infty$, the peaks collapse to a single peak, however for any finite chain, there will be multiple peaks. In fact even in the $N \to \infty$ limit,
 the quasicritical exponents (not the exponents associated with the MCP) will appear in the scaling relation (\ref{scalingGPN}) as also observed  in the  scaling behavior   of defect density following a 
slow quench \ct{dutta10} and that of the fidelity susceptibility \ct{mukherjee10}.
 The analysis of Eq.~(\ref{spectrum}) also shows that no such quasi-critical point exists
      on the paramagnetic side and hence no peak in $d \beta_g/d \gamma$ is expected. Similarly, for $\alpha>2$ the minimum 
      in energy gap occurs right at the MCP where the GP and its derivative vanishes and hence no peak appears as shown in Fig.~(\ref{Fig:qcpnqcp2}). 
For $\alpha = 2$ only one peak of $d \beta_g/d \gamma$ is observed close to the MCP  though the height of the peak saturates for large
$N$ to a constant 
value as the exponent  $(4/\alpha - 2)$ in Eq.~(\ref{scalingGPN}) vanishes in the limit  at $\alpha \to 2$ (Fig.~(\ref{Fig:a2})).
Investigating figures (\ref{Fig:qcpnqcp2}) and (\ref{Fig:a2}) closely,
 we observed peaks at the Ising critical points $h=-1$; this peak appears for all values of $\alpha$ and scales 
 logarithmically with $N$ yielding $\nu=1$ as reported earlier \ct{zhu06}. 

 From Eqs.~(\ref{h2})-(\ref{spectrum}), one finds that  for a finite chain with isotropic interaction ($\gamma=0$), $\theta_{\phi} = 0$ or $\pi$ so that $\beta_g=0$ or $2 \pi$
for $h > 1$ whereas  for  $h \leq 1$, $\beta_g= 2 \pi - 2 \arccos(h)$ \ct{zhu06,zhu08}.  Additionally, in the thermodynamic limit, for infinitesimally small $\gamma$,  
we can always find a solution  $\phi_0$  such that $\cos \phi_0 + (1 - |\gamma|^{\alpha} sgn(\gamma)) = 0 $ but  
$ \Lambda_{\phi_0}= \gamma \sqrt{1-(1 - |\gamma|^{\alpha} sgn(\gamma))^2} \neq 0$ (see Eq.~(\ref{costheta})). This leads to
 $\theta_{\phi_0}=\pi /2$ and  and a non-trivial geometric phase $\beta_g= \pi$ for $\gamma \to 0$. 
 
The relation between the geometric phase and the transverse magnetization ($M_z$) at zero temperature given by the relation $ \beta_g = \pi + \pi M_z$,  has been established close to the Ising
transition\ct{quan09}. Close to the MCP `A', if one defines an $\alpha$-dependent magnetization $M(\alpha)$ through the derivative $\partial \Lambda_{\phi}/\partial h$ for a given path (see 
Eq.~(\ref{path})), it can be shown that $M(\alpha) = M_z + \partial \Lambda_{\phi}/\partial \gamma (\partial \gamma/\partial h)(= 1/\Lambda_{\phi} [(h+ \cos k) - (2/\alpha)\gamma^{2-\alpha} \sin^2 k])$, where the transverse magnetization $M_z$ is related
to the geometric phase as discussed in the reference \ct{quan09}. One can therefore conclude that  close to a MCP the derivative of the geometric phase along a path can not be
related to the transverse magnetic susceptibility directly.


\section{Conclusion}

In this work, the scaling properties of GP have been studied close to a quantum MCP of a spin-1/2 transverse $XY$ spin chain. We show that the scaling
of the derivative of the GP depends on the path of approaching the MCP. It shows a peak whenever the system hits a `quasi-critical point' leading
to a series of peaks on the ferromagnetic side of the MCP.
The peak diverges with the system size with an exponent that varies continuously with
path up to a limiting path given by $\alpha = \alpha_c =2$ where we observe a saturation with $N$.  On the paramagnetic side and also for paths beyond the limiting
path (i.e., for $\alpha >2$), the system does not encounter any quasi-critical point causing the peaks in the derivative to disappear. 

We note that in a recent study,
 a similar behavior of the fidelity susceptibility, namely, occurrence of series of peaks on the ferromagnetic side of the MCP has been reported \ct{mukherjee10}. The  fidelity susceptibility 
has also been found to diverge with the system size with an exponent that continuously varies with the path up to $\alpha=\alpha_c=2$. Our study therefore points to a 
deep connection
between the scaling of the fidelity susceptibility and GP close to quantum critical as well as multi-critical points, as was predicted in the study of quantum geometric tensor 
\ct{venuti07, giorda07}. Questions may remain on the nature of the quasi-critical points, 
especially whether
they are generic or specific to this particular spin chain.  These points may appear in the present model due to the existence of the line of anisotropic critical 
points
with continuously varying ordering wave vectors. However, recent studies including the present work reveal that in all dynamical responses near a quantum MCP, e.g., defect generation following
slow and rapid quenches \ct{deng09, dutta10}, fidelity susceptibility \ct{mukherjee10} as well as in the scaling of the GP, these quasicritical points play a dominant role
 depending on the path of approach.
 
Clearly, the scaling presented in Eq.~(\ref{scalingGPN}) and the limiting value $\alpha=\alpha_c(=2)$ are established close to a MCP in a transverse XY 
chain. For a generic MCP, the scaling law and the value of $\alpha_c$ or the position of the peaks are expected to be different as they depend on the spectrum; however near a generic MCP if quasicritical points
 exist, the associated exponents shall dictate the equivalent scaling relations and the limiting path, and the derivative of the GP will show peaks at the quasicritical points.

\begin{center}
{\bf Acknowledgements}
\end{center}

AP and AD acknowledge CSIR, New Delhi, India, for financial support. We acknowledge Debanjan Chowdhury and Anatoli Polkovnikov for discussions on related issues.

\end{document}